%% file: main.tex
\documentclass[a4paper,11pt]{article}
\pdfoutput=1 % if your are submitting a pdflatex (i.e. if you have
\usepackage{jcappub} % for details on the use of the package, please
\usepackage[T1]{fontenc} % if needed

\input{settings/packages}

\input{settings/newcommands}

\title{\LARGE Precision Gamma-Ray Constraints for Sub-GeV Dark Matter Models}

\author[a]{Adam Coogan,}
\author[b,c]{Logan Morrison,}
\author[b,c]{and Stefano Profumo}

\affiliation[a]{GRAPPA, Institute of Physics, University of Amsterdam, 1098 XH Amsterdam, The Netherlands}
\affiliation[b]{Department of Physics, University of California, 1156 High St., Santa Cruz, CA 95064, USA}
\affiliation[c]{Santa Cruz Institute for Particle Physics, Santa Cruz, CA 95064, USA}

\emailAdd{a.m.coogan@uva.nl}
\emailAdd{loanmorr@ucsc.edu}
\emailAdd{profumo@ucsc.edu}

\abstract{The indirect detection of dark matter particles with mass below the GeV scale has recently received significant attention. Future space-borne gamma-ray telescopes, including All-Sky-ASTROGAM, AMEGO, and GECCO, will probe the MeV gamma-ray sky with unprecedented precision, offering an exciting test of particle dark matter in the MeV-GeV mass range. While it is typically assumed that dark matter annihilates into only one Standard Model final state, this is not the case for realistic dark matter models. In this work we analyze existing indirect detection constraints and the discovery reach of future detectors for the well-motivated Higgs and vector-portal models using our publicly-available code \texttt{Hazma}. In particular, we show how to leverage chiral perturbation theory to compute the dark matter self-annihilation cross sections into final states containing mesons, the strongly-interacting Standard Model dynamical degrees of freedom below the GeV scale. We find that future telescopes could probe dark matter self-annihilation cross sections orders of magnitude smaller than those presently constrained by cosmic microwave background, gamma-ray and terrestrial observations.}

\begin{document}
\maketitle
\flushbottom

\section{Introduction}
\label{sec:introduction}
\input{sections/introduction}

\section{Theoretical Framework and its Validity}%
\label{sec:theoretical_framework}
\input{sections/theoretical_framework}

\section{Searching for MeV Dark Matter with Future Gamma Ray Telescopes}
\label{sec:experimental_reach}
\input{sections/experimental_reach}

\section{Conclusions}\label{sec:conclusion}
\input{sections/conclusions}

\acknowledgments
LM and SP are partly supported by the U.S.\ Department of Energy grant number de-sc0010107. We acknowledge with gratitude early collaboration on this project with Francesco D'Eramo. Francesco D'Eramo was the first to put forward the idea of using chiral perturbation theory in the context of accurately calculating the annihilation and decay products of MeV dark matter. He was also responsible for several of the early calculations, part of the writing of the text, and for overall educating us all on numerous topics in chiral perturbation theory. We thank Michael Peskin for important input and feedback, and Regina Caputo for discussions about MeV gamma-ray observatories. Finally, we thank Eulogio Oset and Jose Antonio Oller for clarifying some points about unitarized chiral perturbation theory.

\bibliographystyle{apsrev}
\bibliography{refs}

\end{document}

%% file: settings/packages.tex
\usepackage{lipsum,multirow,slashed,comment,feynmp}

\usepackage{xargs} % Use more than one optional parameter in a new commands
\usepackage[pdftex, dvipsnames]{xcolor}  % Coloured text etc.
\usepackage{cleveref}
\usepackage[utf8x]{inputenc}

\usepackage{amsmath}
\usepackage{amssymb}
\usepackage{amsthm}
\usepackage{amsfonts}
\usepackage{verbatim}
\usepackage{enumerate}
\usepackage{accents}

\usepackage{physics}
\usepackage{tensor}
\usepackage{siunitx}
\usepackage{wasysym}

\usepackage{graphicx}
\usepackage{booktabs}

%% file: settings/newcommands.tex
\newcommand{\chipt}{\mathrm{ChPT}}

\renewcommand\d{\partial}

\renewcommand\u[1]{\text{#1}} % No space
\renewcommand\L{\mathcal{L}}

\renewcommand\d{\partial}

\DeclareSIUnit\year{yr}

\newcommandx{\unsure}[2][1=]{\todo[linecolor=red,backgroundcolor=red!25,bordercolor=red,#1]{#2}}
\newcommandx{\change}[2][1=]{\todo[linecolor=blue,backgroundcolor=blue!25,bordercolor=blue,#1]{#2}}
\newcommandx{\info}[2][1=]{\todo[linecolor=OliveGreen,backgroundcolor=OliveGreen!25,bordercolor=OliveGreen,#1]{#2}}
\newcommandx{\improvement}[2][1=]{\todo[linecolor=Plum,backgroundcolor=Plum!25,bordercolor=Plum,#1]{#2}}
\newcommandx{\thiswillnotshow}[2][1=]{\todo[disable,#1]{#2}}

\newcommand{\revised}[1]{{#1}}

%% file: sections/introduction.tex
The paradigm of Weakly-Interacting Massive Particles (WIMPs), a class of  dark matter particle candidates with weak-scale mass and charged under Standard Model weak interactions, is extraordinarily compelling. WIMPs are found in myriad extensions to the Standard Model of particle physics, their relic abundance from the Early Universe is often very close to the observed abundance of dark matter in the Universe, and the fact that they share weak interactions with the Standard Model makes them in principle discoverable through a broad array of experimental techniques (for a recent review see e.g. Ref.~\cite{Arcadi:2017kky}).

In the last few decades, the ``natural'' scales for the cross sections relevant to the detection of WIMP dark matter have been targeted by numerous experimental searches. In particular, direct dark matter detection -- the search for the minuscule energy deposition that a WIMP would impart to a nucleus in a low-background detector -- has ruled out large swaths of parameter space for WIMPs that interact through, for example, the exchange of Standard Model $Z$ bosons or Higgs bosons, in a broad WIMP mass range between a few GeV and up to several TeV. WIMPs have failed to appear at colliders, where they are searched in the form of missing energy/momentum. Finally, the pair-annihilation cross section expected for a thermal relic from the early universe and weak-scale mass has been extensively searched for with gamma-ray as well as cosmic-ray space-borne experiments, generally with null results (though some controversial possible ``detections'' are still debated).

The WIMP paradigm for thermal production in the early universe -- the notion that the particle species making up the dark matter was once in thermal equilibrium with Standard Model particles, subsequently falling out of chemical equilibrium and ``freezing out'' with the right relic density -- is actually not unique to the weak scale and to weak interactions (see e.g. the ``WIMP-less miracle'' described in Ref.~\cite{Feng:2008ya}). Assuming the existence of new force mediators, lighter particles, much below the Lee-Weinberg limit~\cite{Lee:1977ua} (which states that particles with neutrino-like weak interactions would be produced with an excessively large density if their mass were much lighter than a few GeV) provide perfect WIMP-like dark matter candidates from the standpoint of production, as well as for that of possible detection.

An interesting range for the mass of such light, WIMP-like dark matter candidates is the MeV scale. The direct detection of MeV dark matter is challenging, since the recoil energy is well below the threshold sensitivity of most current detectors. New ideas on how to experimentally search for MeV dark matter scattering have however been investigated~\cite{Knapen:2017xzo}. This area, as well as the related question of how other cosmological and collider constraints circumscribe the target parameter space, continues to witness intense activity (see e.g.~\cite{Krnjaic:2015mbs,Dolan:2014ska,Knapen:2017xzo, Lehmann:2020lcv}).

The indirect detection of MeV-scale WIMP-like dark matter particles  --  the detection of the debris of dark matter annihilation or decay -- is an especially promising and timely arena. On the one hand, new telescopes will quite literally revolutionize the relevant energy range for indirect MeV dark matter detection (as well as for other searches for new physics, such as radiation from the evaporation of light primordial black holes, see e.g.~\cite{Coogan:2020tuf}), as we describe below. On the other hand, from a theoretical standpoint, the lack of specific observational facilities has been responsible for somewhat of a gap in the understanding of the details of the photon spectrum to be expected from the annihilation or decay of MeV dark matter, especially compared to what is known and established for ``traditional'' WIMP candidates.

\begin{figure}
    \centering
    \includegraphics[width=\linewidth]{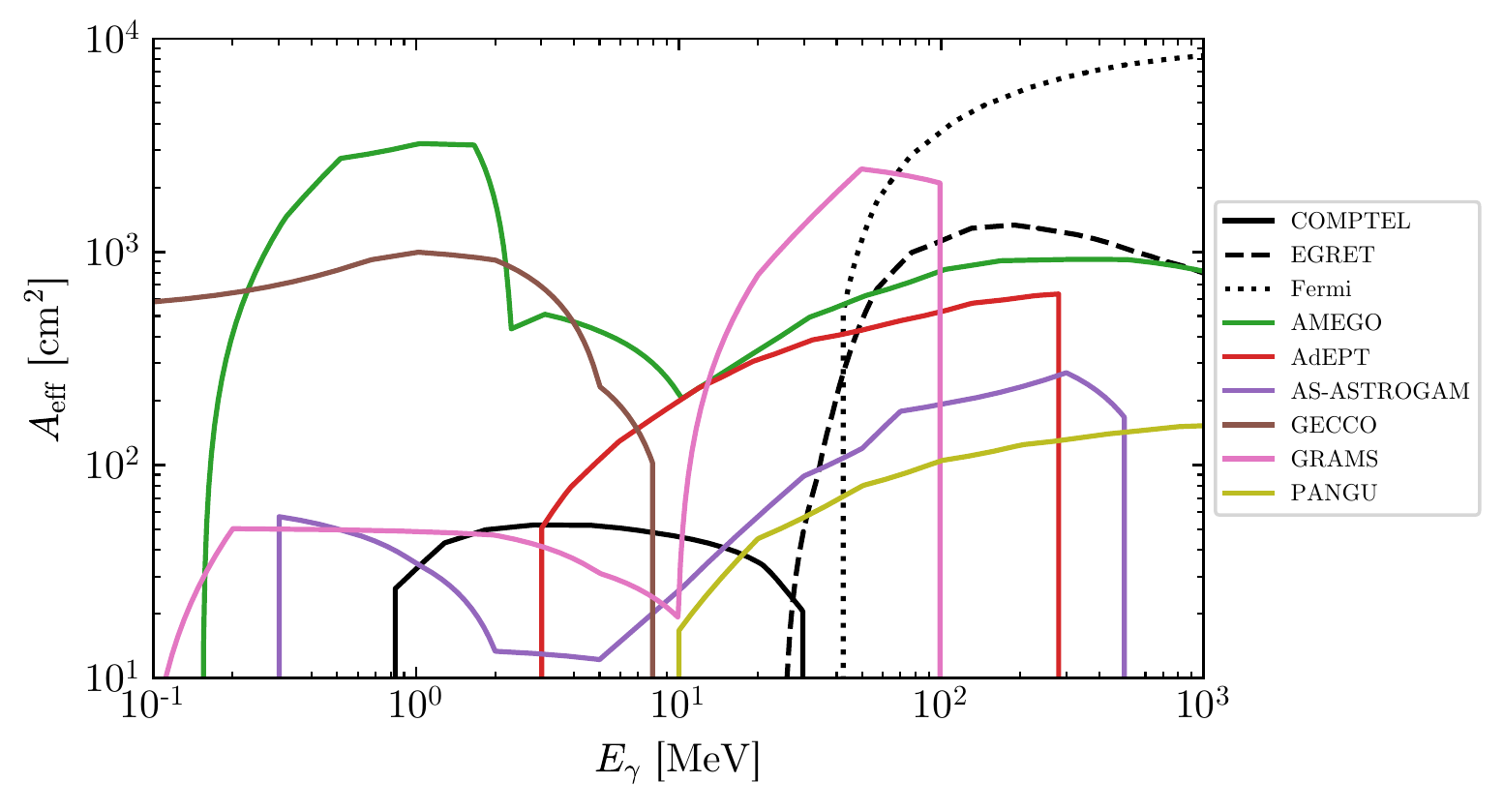}
    \caption{The effective area as a function of energy for past, existing, and planned telescopes under consideration in our study. The effective area of MAST is not shown since it is significantly larger than for other telescopes ($\SIrange{6.5e4}{1.2e5}{\cm^2}$ for $E_\gamma = \SIrange{10}{1000}{\mega\eV}$).}
    \label{fig:aeff}
\end{figure}

As with WIMPs, the annihilation of MeV-scale dark matter can produce identifiable and sometimes unmistakable features in the electromagnetic spectrum. Gamma rays from WIMPs in the \SIrange{0.5}{250}{\mega\eV} mass range would lie predominantly in the range $\order{\SIrange{0.1}{100}{\mega\eV}}$, which includes the prominent neutral pion decay peak centered at $\sim \SI{70}{\mega\eV}$. This energy window was last explored by COMPTEL~\cite{Strong:1998ck} and EGRET~\cite{Hunger:1997we} in the 1990s, leading to a couple order-of-magnitude gap (in terms of spectral energy density point source sensitivity) relative to the $\gtrsim \SI{1}{\giga\eV}$ and x-ray portions of the electromagnetic spectrum (see \cref{fig:aeff}). Several experiments have been proposed to close this gap: GAMMA-400~\cite{Galper:2014pua,Egorov:2020cmx}, Advanced Compton Telescope (ACT)~\cite{Boggs:2006mh}, Advanced Energetic Pair Telescope (AdEPT)~\cite{Hunter:2013wla}, PANGU~\cite{pangu,pangu_aeff}, GRAMS~\cite{grams,grams_loi}, MAST~\cite{mast}, AMEGO~\cite{amego,McEnery:2019tcm,Kierans:2021kpg}\footnote{See also \url{https://asd.gsfc.nasa.gov/amego/}} and All-Sky-ASTROGAM~\cite{as_astrogam} (a scaled-back version of e-ASTROGAM~\cite{eASTROGAMWhitebook}).\footnote{
    We do not project discovery reach for ACT since the energy-dependent effective area is not available in the literature. We also do not consider GAMMA-400 further since its effective area is significantly lower than Fermi's, and a previous study already assessed its discovery reach for dark matter annihilating into photon pairs~\cite{Egorov:2020cmx}, finding it to be a factor of $\sim 1.5$ times more sensitive than Fermi for 2 years of observing time.
} Another promising telescope is the Galactic Explorer with a Coded Aperture Mask Compton Telescope (GECCO), which encapsulates at once the principles of a Compton telescope and of a coded-aperture mask telescope~\cite{gecco_loi}. GECCO's performance in the search for dark matter and new physics was recently explored in Ref.~\cite{Coogan:2021rez}. We show the approximate anticipated effective areas of the various telescopes under consideration here in \cref{fig:aeff}.

The calculation of indirect \revised{gamma-ray} constraints on models of sub-GeV dark matter presents several important technical challenges \revised{that have not been simultaneously addressed in previous studies~\cite{Gonzalez-Morales:2017jkx,Bartels2017,Essig:2013goa,Boddy2015,Kumar:2018heq,Cirelli:2020bpc}}. Firstly, in realistic dark matter models the dark matter typically annihilates into more than one final state. In this case, unlike what holds for mass scales of $\sim \SI{10}{\giga\eV}$ or more, where perturbative QCD is an adequate theoretical tool, since QCD confines below $\Lambda_\mathrm{QCD} \sim \SI{1}{\giga\eV}$ the relevant final-state degrees of freedom for MeV dark matter are light mesons. Annihilation branching fractions in such models can be computed using \emph{chiral perturbation theory} (ChPT), the effective field theory of mesons. A second issue concerns the spectrum of final-state radiation off of charged annihilation final states (including pions, muons and electrons). This is often handled using the Altarelli-Parisi approximation~\cite{Altarelli:1977zs}, but in actuality depends on the spin of the mediator and radiating particles. Lastly, computing photon spectra for models that annihilate into charged pions requires assessing that particle's complex radiative decay chain.

\emph{In this work}, we derive effective Lagrangians describing the interactions of sub-GeV dark matter with light mesons. In particular, we focus on two well-motivated dark matter models (containing mediators that mix with the Higgs and photon) and explain how to match the mediator's interactions with quarks and gluons onto interactions with mesons using ChPT. We then use these Lagrangians and the final-state radiation and radiative decay spectra computed in our previous paper~\cite{hazma} to find the gamma-ray constraints on these models and project which annihilation cross sections could be probed by the aforementioned planned telescopes. Additionally we explore how these complement cosmic microwave background and other constraints. This study thus acts as a companion to our previous work~\cite{hazma}, wherein we computed annihilation spectra and presented plots of the annihilation branching fractions for these models (figs. 3 and 4), but did not explain the ChPT matching procedure or derive detailed model constraints/projections.

The reminder of this paper is organized as follows. In \cref{sec:theoretical_framework} we provide an overview of the parton-level Lagrangian for the models we consider and a discussion of the applicability and validity of chiral perturbation theory for given dark matter and mediator masses. We then describe the specific microscopic models we consider, and the matching between parton-level and hadronic-level Lagrangians. \Cref{sec:experimental_reach} assesses the discovery reach for planned MeV gamma-ray telescopes, and compares with bounds from existing gamma-ray and cosmic microwave background data as well as non-astrophysical observations. Finally, \cref{sec:conclusion} concludes.

%% file: sections/theoretical_framework.tex
Our analysis considers extensions of the Standard Model (SM) defined in terms of microscopic Lagrangians just above the confinement scale $\Lambda ~ \SI{1}{\giga\eV}$. We posit that there is a dark matter particle $\chi$ with sub-GeV mass and a mediator that couples the dark and visible sectors.  At the GeV scale the relevant fundamental SM degrees of freedom are the photon, light leptons $(e, \mu)$, gluons and light quarks $(u, d, s)$. Letting $M$ represent the mediator, the Lagrangians for such models takes the form
\begin{equation}
    \L = \L_\mathrm{SM} + \L_{\rm DM} + \L_{\rm M} + \L_{\rm Int(M)}, \label{eq:generic-lagrangian}
\end{equation}
where the second and third pieces are the free terms for the DM and mediator and the fourth contains mediator-SM and mediator-DM interaction terms. We assume that $\chi$ is a (Dirac) fermion and a SM gauge singlet, so that it cannot couple directly to SM fields in a renormalizable manner.

At the energy scale for annihilations of such DM particles, quarks and gluons are not the right strongly-interacting dynamical degrees of freedom. Instead, we must match \cref{eq:generic-lagrangian} onto one defined in terms of light \emph{mesons}. In the rest of this section, we describe the framework for performing this matching, its range of validity, the particular DM models we consider and how to perform their matching in detail.

\subsection{Low-energy QCD degrees of freedom}

Since the strong interaction is confining below $\sim \SI{1}{\giga\eV}$, the relevant degrees of freedom in the MeV range are pions and other mesons. As was know since about 1980, these degrees of freedom can be described by an effective field theory (EFT) called \emph{chiral perturbation theory} ($\chipt$), which is derived by treating the pions as pseudo-Goldstone bosons under the chiral symmetry group $\operatorname{SU}(3)_L \times \operatorname{SU}(3)_R$~\cite{GASSER1985465,Scherer2003,WEINBERG1979327}. The expansion parameter of the EFT is $p/\Lambda_{\chipt}$, where $p$ is the characteristic momentum scale of the process in question and $\Lambda_{\chipt} \sim 4\pi f_{\pi} \sim \SI{1}{\giga\eV}$ is the cut-off scale for $\chipt$; $f_{\pi} \approx \SI{92}{\mega\eV}$ is called the pion decay constant. At a fixed order in the EFT expansion parameter, we restrict the number of derivatives on meson fields since $\partial_{\mu} \sim p_{\mu}$. Thus, at leading order (i.e. at order $(p/\Lambda_{\chipt})^2$), the most general $\chipt$ Lagrangian density consistent with chiral symmetry is:
\begin{align}
    \mathcal{L}_{\chipt} &= \frac{f^2_{\pi}}{4}\Tr[\qty(D_{\mu}\Sigma)\qty(D^{\mu}\Sigma)^{\dagger}] + \frac{f^2_{\pi}}{4}\Tr[\chi^{\dagger}\Sigma+\Sigma^{\dagger}\chi].
\end{align}
There are a number of items in this Lagrangian density that need explaining: First, $\Sigma$ is the Goldstone matrix transforming under $(L,R)\in\mathrm{SU}(3)_{\mathrm{L}}\times\mathrm{SU}(3)_{\mathrm{R}}$ as $\Sigma\to R\Sigma L^{\dagger}$. Explicitly, $\Sigma$ is given by the exponential of a $3\times3$ matrix containing pions, kaons and the $\eta$:\footnote{Note that what we refer to as the $\eta$ is technically the $\eta_8$. The physical $\eta$ is instead a mixture of $\eta_8$ and $\eta_1$, the field associated with the (anomalous) $\mathrm{U}_A(1)$ symmetry. However, since $\eta = \eta_8 \, \cos\theta - \eta_1 \, \sin\theta$ and $\theta \approx -\ang{11}$ we ignore this subtlety~\cite{Zyla:2020zbs}.} $\Sigma = \exp(i\Phi/f_{\pi})$, with $\Phi$ given by
\begin{align}
    \Phi = \mqty(
       \pi^{0} + \frac{1}{\sqrt{3}}\eta & \sqrt{2}\pi^{+} & \sqrt{2}K^{+}\\
       \sqrt{2}\pi^{-} & -\pi^{0} + \frac{1}{\sqrt{3}}\eta & \sqrt{2}K^{0}\\
       \sqrt{2}K^{-} & \sqrt{2}\bar{K}^{0} & -\frac{2}{\sqrt{3}}\eta
    )
\end{align}
The covariant derivative acting on the Goldstone matrix, $D_{\mu}$, contains the $\mathrm{SU}(3)_{\mathrm{L}}$ and $\mathrm{SU}(3)_{\mathrm{R}}$ up, down and strange quark currents gathered into $3\times3$ matrices $\ell_{\mu}$ and $r_{\mu}$. These are taken to be $\order{p}$. The chiral currents transform under $(L,R)\in \mathrm{SU}(3)_{\mathrm{L}}\times\mathrm{SU}(3)_{\mathrm{R}}$ as $r_{\mu}\to Rr_{\mu}R^{\dagger}$ and $\ell_{\mu}\to L\ell_{\mu}L^{\dagger}$. These transformation rules restrict the form of the covariant derivative to
\begin{align}
    D_{\mu}\Sigma = \partial_{\mu}\Sigma -i r_{\mu}\Sigma + i\Sigma\ell_{\mu},
\end{align}
which results in an object that transforms as $\Sigma$ under chiral transformations: $D_{\mu}\Sigma \to R\qty(D_{\mu}\Sigma)L^{\dagger}$. The second term in the leading order chiral Lagrangian encodes the masses of the mesons. The $\chi$ field is taken to be a spurion field responsible for chiral symmetry breaking. The spurion field transforms in the same way $\Sigma$ does (namely $\chi\to R\chi L^{\dagger}$) and is counted as $\order{p}$. The expansion of $\chi$ around its vacuum expectation value is
\begin{align}
    \chi = 2B_{0}(M_{q} + s + i p),
\end{align}
where $s$ and $p$ are scalar and pseudoscalar up, down and strange quark currents and $M_{q}$ is the quark mass matrix. With no external fields, the quark mass matrix breaks chiral symmetry. The constant $B_{0}$ is related to the expectation value of the quark condensate via $B_{0} = -\expval{\bar{q}q}/3f_{\pi} \sim \SI{2.3}{\giga\eV}$.

\subsection{ChPT Applicability}

Before moving on to describing the dark matter and mediator interactions with mesons, we would like to briefly explain the regime of validity of chiral perturbation theory. As the magnitude of the meson momentum $p$ becomes comparable to $\Lambda_{\rm ChPT}$, loop diagrams from the leading-order Lagrangian and tree diagrams from the next-to-leading order Lagrangian contribute comparably to tree-level diagrams from the leading order Lagrangian, signaling a breakdown in the effective theory.

However, the convergence of ChPT is actually disrupted at a lower scale than $\Lambda_{\rm ChPT}$ by hadronic resonances such as the scalar $f_0(500)$~\cite{Pelaez:2015qba} and the vector meson $\rho$. These induce significant interactions between hadrons produced by DM annihilation or mediator decay at center-of-mass energies of $p \gtrsim \SI{500}{\mega\eV}$. The $\rho$ and other vector resonances can be included in extensions of chiral perturbation theory~\cite{Ecker:1988te}. The $f_0(500)$ cannot be incorporated in this framework, but can instead be accounted for using chiral unitary techniques in meson-meson scattering~\cite{Pelaez:2015qba}, which can be extended to compute corrections to cross sections for DM annihilation processes like $\text{DM DM} \to \pi \pi$. However, determining the impact on final state radiation processes like $\text{DM DM} \to \pi \pi \gamma$ is much more technically challenging. We leave this for future work, restricting the present analysis to the narrow but well-controlled $p \leq \SI{500}{\mega\eV}$ part of parameter space.

These considerations imply that the analysis in Ref.~\cite{Kumar:2018heq} pushes leading-order ChPT beyond its range of validity. That study focuses on final states containing an $\eta$ meson and a $\pi^0$ or $\pi^+ \pi^-$. The center of mass energies for these final states with all particles produced at rest are $p_{\pi^0 \eta} \sim \SI{680}{\mega\eV}$ and $p_{\pi^+ \pi^- \eta} \sim p_{\pi^0 \pi^0 \eta} \sim \SI{825}{\mega\eV}$, well beyond the scale at which resonances must be taken into account. For example, the strength of final state interactions for $\mathrm{DM~DM} \to \pi^0 \eta$ is related to the elastic $\pi^0 \eta \to \pi^0 \eta$ scattering cross section by the Watson final state theorem~\cite{PhysRev.95.228}. This amplitude is dominated by exchange of the $f_0(500)$ and $a_0(980)$ scalar resonances~\cite{Black:1999dx}, with little contribution from leading-order ChPT.~\footnote{Note that the $f_0(500)$ is instead denoted by $\sigma(560)$ in Ref.~\cite{Black:1999dx}.} As another example, the final state interactions relevant for $\mathrm{DM~DM} \to \pi \pi \eta$ with $J^{PC}= 0^{-+}$ are the same as for the process $\eta' \to \pi \pi \eta$. The value for the branching fraction computed with leading-order ChPT is only 3\% of the measured value~\cite{Escribano:2010wt}. Successfully predicting the experimentally-measured Dalitz plot parameters requires combining next-to-leading order large-$N_c$ ChPT with a unitarization procedure to account for final state interactions.

\begin{figure}
    \centering
    \includegraphics[width=\linewidth]{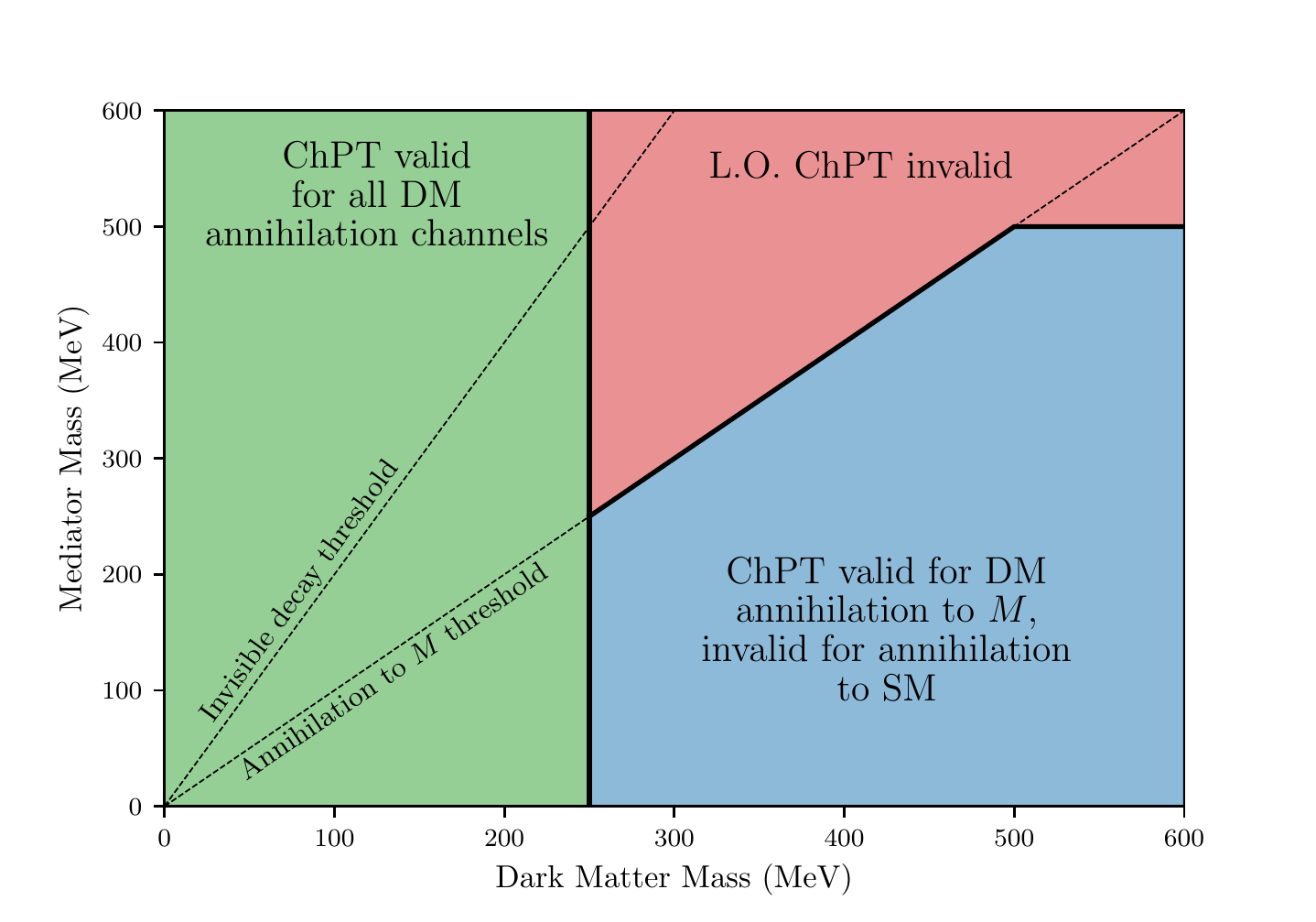}
    \caption{Illustration of DM masses $m_{\rm DM}$ and mediator masses $m_M$ for which the leading-order chiral perturbation theory calculations in this paper can be considered reliable.}
    \label{fig:phase_diagram}
\end{figure}

The diagram in \cref{fig:phase_diagram} shows in detail the range of validity of this work, with the red region indicating where leading order ChPT calculations cannot be trusted due to resonances and/or convergence issues with ChPT. The dashed lines indicate parameter space for which the mediator can decay invisibly and the DM can annihilate into the mediator. In the green region, the DM annihilation cross sections are well-described by leading-order  ChPT. In the blue one, ChPT cannot be used to compute the cross section for annihilation into SM final states. However, annihilation of the dark matter into mediators followed by their subsequent decay can be accurately treated using ChPT in this region.

As a consequence of this analysis, we neglect annihilation and mediator decays into strange mesons $K^+$, $K^0$ and $\eta$. Producing kaons in appreciable numbers in DM annihilations requires $m_K < m_{\rm DM} < m_M$; alternatively, producing them through annihilation into mediators followed by mediator decay requires $2 m_K < m_M < m_{\rm DM}$. Both these parts of parameter space are within the red region of Fig.~\ref{fig:phase_diagram}. The $\eta$ mass, $m_\eta = \SI{547}{\mega\eV}$, is also outside the range of validity for a leading order ChPT analysis, and will be neglected throughout. The strange quark and associated couplings will thus be ignored in this work, though they are an important input for any chiral perturbation theory analysis beyond leading order.

\subsection{Microscopic Models}

We consider two choices for the mediators, a (real) scalar mediator $S$ and a vector mediator $V$. The free Lagrangian for the scalar mediator reads
\begin{equation}
    \L_S = \frac{1}{2} (\d_\mu S)(\d^\mu S) - \frac{1}{2} m_S^2 S^2,
\end{equation}
with the following interaction Lagrangian:
\begin{equation}
    \L_{\rm Int(S)} = -S \left( g_{S\chi} \bar{\chi} \chi + \frac{g_{Sf}}{\sqrt{2}} \sum_{f} y_f  \bar{f} f \right) + 
    \frac{S}{\Lambda} \left(g_{SF} \frac{\alpha_{\rm EM}}{4\pi} F_{\mu\nu} F^{\mu\nu} + g_{SG} \frac{\alpha_s}{4\pi} G_{\mu\nu}^a G^{a \mu\nu} \right).
    \label{eq:LagScalar}
\end{equation}
In the expression above, we assume that the interactions of the scalar $S$ with SM fermions is mediated by Yukawa interactions, where $y_f = \sqrt{2} m_f / v_h$ and $v_h = 246$ GeV is the vacuum expectation value of the SM Higgs (this choice is motivated by the framework of minimal flavor violation \cite{Buras:2000dm,DAmbrosio:2002vsn}). We also included non-renormalizable operators describing effective interactions of the mediator $S$ with the low-energy gauge boson degrees of freedom, photons and gluons; these interactions are invariably generated by integrating out heavier degrees of freedom including those responsible for ensuring gauge invariance of the theory at high energy. 

In what follows, we fix the parameters $g_{Sf}$, $g_{SF}$ and $g_{SG}$ assuming the Higgs portal scenario, where a quartic coupling $\lambda H^\dagger H S^2$ induces a mixing of the mediator $S$ with the SM Higgs boson. In this case, indicating with $\theta$ the relevant mixing angle, $g_{Sf}=-\sin\theta$, $g_{SF}=5\sin\theta/6$ and $g_{SG}=-3\sin\theta$ (see \cite{Marciano:2011gm} for details), and the effective mass scale $\Lambda=v_h$.

In the case of a vector mediator, the free mediator Lagrangian reads
\begin{equation}
    \L_V =  - \frac{1}{4} V_{\mu\nu} V^{\mu\nu} + \frac{1}{2} m_V^2 V_\mu V^\mu,
\end{equation}
with interaction terms
\begin{equation}
    \L_{\rm Int(V)} =  V_\mu \left( g_{V \chi} \bar{\chi} \gamma^\mu \chi + \sum_{f} g_{Vf} \bar{f} \gamma^\mu f \right)  - \frac{\epsilon}{2} V^{\mu\nu} F_{\mu\nu}  \ . \label{eq:LagVector}
\end{equation}
In the expression above, the sum over fermions includes all light degrees of freedom, $f \in \{e,\mu,u,d,s\}$, while $\epsilon$ is a kinetic mixing parameter. Notice that upon $A_\mu \rightarrow A_\mu - \epsilon V_\mu$ the kinetic mixing term disappears, and the charge assignments in turn change, e.g. $g_{Vf} \; \rightarrow g_{Vf} - \epsilon e Q_f$, where $Q_f$ is the electric charge of fermion $f$. In what follows, we posit that the interactions with SM fermions be only originating from said kinetic mixing, and thus $g_{Vf}=-\epsilon eQ_f$. 

\subsection{Matching}

In this section we demonstrate the procedure of matching various terms from the Lagrangians defined at energies above $\Lambda_{\mathrm{QCD}}$ onto the chiral Lagrangian. In particular, we will be interested in matching the following terms:
\begin{align}
    &S\bar{q} \vb{G}_{Sq}q, & 
    &SG^a_{\mu\nu}G^{\mu\nu,a}, & 
    &V_{\mu}\bar{q}\qty(\vb{G}_{Vq})\gamma^{\mu}q
\end{align}
where \(S\) and \(V_{\mu}\) are the scalar and vector mediators, \(q = \mqty(u &d &s)\) and \(\vb{G}_{Sq},\vb{G}_{Vq}\) are \(3\times3\) coupling matrices. Below, we provide the details for matching each of the above terms onto the Chiral Lagrangian. 

\subsubsection*{\boldmath\texorpdfstring{$S\bar{q}G_{Sq}q$}{S qbar GSq q}}

The term $S\bar{q}G_{Sq}q$ closely resembles the quark mass term: $\bar{q}M_q q$. In fact, we match the former and latter terms onto the chiral Lagrangian using the same technique. Recall that the quark mass terms are matched onto the chiral Lagrangian by introducing a spurion field $\chi$ transforming as $\chi\to R\chi L^{\dagger}$ under chiral transformations. Then, we form a chiral invariant using $\Tr[\chi\Sigma^\dagger + \mathrm{h.c.}]$. $\chi$ then is used to break chiral symmetry by setting it to its vacuum state $\chi = 2BM_{q}$. Thus, the mass term is matched as follows:
\begin{align}
    -\bar{q}M_q q &\to \dfrac{f_{\pi}^2}{4}\Tr[\chi\Sigma^{\dagger}+\mathrm{h.c.}], &
    \chi = 2B M_{q}.
\end{align}
Generalizing by adding interaction terms of the form $S\bar{q}G_{Sq}q$, with $G$ a $3\times3$ matrix, we change $\chi$ to:
\begin{align}
    \chi \to  2B\qty(M_q + S G_{Sq})
\end{align}
In the Higgs portal model, we have scalar Yukawa interactions with $G_{Sq}=-g_{Sf}M_{q}/v_{h}$. These, together with the quark mass terms, are therefore matched as:
\begin{align}
    -\sum_{i=3}^{3}\qty(1+g_{Sf}\frac{S}{v_{h}})m_{q}\bar{q}_iq_i &\to \dfrac{f_{\pi}^2}{4}\Tr[\chi\Sigma^{\dagger}+\mathrm{h.c.}], & \chi = 2BM_q\qty(1+g_{Sf}\frac{S}{v_{h}}).
\end{align}

\subsubsection*{\boldmath\texorpdfstring{$\frac{\alpha_{s}}{4\pi}\frac{S}{\Lambda}G^a_{\mu\nu}G^{\mu\nu,a}$}{αs/4π S/Λ G G}}

Terms of the form $\frac{\alpha_{s}}{4\pi}\frac{S}{\Lambda}G^a_{\mu\nu}G^{\mu\nu,a}$ frequently arise when integrating out heavy quarks. For example, this term arises in the Higgs portal model when integrating out the top, bottom and charm quarks at one-loop. The matching of $SG^a_{\mu\nu}G^{\mu\nu,a}$ onto the chiral Lagrangian is performed by utilizing the trace anomaly~\cite{Donoghue:1990xh,chivukula1989higgs}. The trace anomaly relates the divergence of the dilatation current $\partial^{\mu}d_{\mu}$ to the gluon kinetic term and quark mass operators.
The key insight in matching $\alpha_s G^2$ onto the chiral Lagrangian is that $\partial_\mu d^\mu$ is RGE-invariant~\cite{Grinstein:1988wz}. In Ref.~\cite{chivukula1989higgs}, it was shown that, at leading order in perturbation theory, $\frac{\alpha_{s}}{4\pi}\frac{S}{\Lambda}G^a_{\mu\nu}G^{\mu\nu,a}$ can be written as: 
\begin{align}
    \frac{\alpha_{s}}{4\pi}\frac{S}{\Lambda} G^a_{\mu\nu} G^{\mu\nu,a} = -\frac{2}{\beta_{0}}\frac{S}{\Lambda}\qty[\partial_{\mu}d^{\mu} -\sum_{q=u,d,s}m_{q}\bar{q}q]
\end{align}
where $\beta_{0}=11-\frac{2}{3}N_{f}$ is the leading order $\beta$-function for the QCD coupling constant with $N_{f}$ flavors. Since $\partial_{\mu}d^{\mu}$ is RGE-invariant, it is matched onto the chiral Lagrangian by simply computing the divergence of the scale current using the chiral Lagrangian:
\begin{align}
    \partial_{\mu}d^{\mu} = -\frac{f_{\pi}^2}{2}\Tr[\qty(D_{\mu}\Sigma)\qty(D_{\mu}\Sigma)^{\dagger}] - 2f_{\pi}^2B_{0}\Tr[M_{q}\qty(\Sigma + \Sigma^{\dagger})]
\end{align}
We can match $m_{q}\bar{q}q$ onto the chiral Lagrangian using techniques described above. The result is:
\begin{align}
    \frac{\alpha_{s}}{4\pi}\frac{S}{\Lambda} G^a_{\mu\nu} G^{\mu\nu,a} &\to \frac{f_{\pi}^2}{\beta_{0}}\frac{S}{\Lambda}\Tr[\qty(D_{\mu}\Sigma)\qty(D_{\mu}\Sigma)^{\dagger}] 
    + \frac{3f_{\pi}^2B_{0}}{\beta_{0}}\frac{S}{\Lambda}\Tr[M_{q}\qty(\Sigma + \Sigma^{\dagger})] + \cdots
\end{align}
where the $(\cdots)$ contains terms with more than one $S$ coming from matching $-2S/(\beta_0\Lambda)m_q \bar{q}q$ and $g_{Sf}(S/v_h)\bar{q}q$ onto the chiral Lagrangian. These are irrelevant for this study so we omit them.

\subsubsection*{\boldmath\texorpdfstring{$V_{\mu}\bar{q}G_{Vq}\gamma^{\mu}q$}{V qbar GVq γμ q}}
External vector currents are matched onto the chiral Lagrangian through the covariant derivative:
\begin{align}
    D_{\mu}X = \partial_{\mu}X -ir_{\mu}X + iX\ell_{\mu}
\end{align}
where $r_{\mu}$ and $\ell_{\mu}$ are right-handed and left-handed currents. Given a quark vector current $V^{\mu}J_{\mu} = G_{Vq}V^{\mu}\bar{q}\gamma^{\mu}q$, we can identify:
\begin{align}
    r_{\mu} = \ell_{\mu} &= G_{Vq}V_{\mu} + \cdots
\end{align}
where the $\cdots$ represent other currents (such as the electromagnetic currents.)

There are additional terms outside of the chiral expansion that are necessary for our studies. Wess, Zumino~\cite{Wess:1971yu} and Witten~\cite{Witten:1983tw} showed the existence of an additional term in the chiral Lagrangian which gives rise to the neutral pion's decay into two photons:
\begin{align}
    \mathcal{L}_{\pi^{0}\gamma\gamma} = -\frac{e^2}{32\pi^2}\epsilon_{\mu\nu\rho\sigma}F^{\mu\nu}F^{\rho\sigma}\dfrac{\pi^0}{f_{\pi}}
\end{align}
where $F_{\mu\nu}$ is the photon field strength tensor and $\epsilon_{\mu\nu\rho\sigma}$ is the 4-dimensional Levi-Civita symbol. In the case of kinetic mixing model, the shift of the photon field $A_{\mu}\to A_{\mu}-\epsilon V_{\mu}$, results in the following vector-photon-pion coupling:
\begin{align}
    \mathcal{L}_{\pi^{0}\gamma\gamma} \to
    -\frac{e^2}{32\pi^2}\epsilon_{\mu\nu\rho\sigma}F^{\mu\nu}F^{\rho\sigma}\dfrac{\pi^0}{f_{\pi}}
    -\frac{\epsilon e^2}{16\pi^2}\epsilon_{\mu\nu\rho\sigma}F^{\mu\nu}V^{\rho\sigma}\dfrac{\pi^0}{f_{\pi}}
\end{align}
Hence, we pick up an additional $V\gamma\pi^0$ coupling which gives important contributions to the dark matter annihilation spectrum when annihilations into two-pion channels are forbidden.

%% file: sections/experimental_reach.tex
We now turn to assessing the prospects for future MeV gamma-ray telescopes to explore the parameter space of dark matter for the Higgs portal and dark photon portal models. After briefly summarizing how to compute the gamma-ray yield from dark matter annihilation, we describe the calculation of constraints from existing gamma-ray telescopes, the cosmic microwave background and other particle physics observations. We then discuss our results for the anticipated performance of future MeV gamma-ray telescopes.

\subsection{Gamma-ray constraints}

Gamma rays can be produced through three mechanisms when dark matter self-annihilates: direction photon production and radiation from and radiative decays of final state particles.\footnote{
    Dark matter annihilations also produce photons indirectly through astrophysical processes involving the annihilation products, such as synchrotron radiation in regions with magnetic fields and inverse-Compton scattering of cosmic microwave background photons. This so-called secondary emission spectrum is challenging to compute accurately given the substantial astrophysical uncertainties involved~\cite{Colafrancesco:2005ji,Colafrancesco:2006he, McDaniel:2017ppt} and we thus neglect it here.
} Photons from the first mechanism are simple to account for, contributing delta function spectra in the cases considered in this work ($\gamma \gamma$ in the Higgs portal model and $\pi^0 \gamma$ for the dark photon model). The spectrum of final-state radiation is dependent on the details of the mediator's couplings to the final-state particle and whether the particle is a scalar or fermions, so its accurate calculation requires the effective Lagrangians provided above. Analytic expressions for these spectra were presented in eqs. 4.8 - 4.11 in our previous work~\cite{hazma}. There we also provide the radiative decay spectra of the neutral pion (eq. 4.20), muon (eq. 4.21) and charged pion (eqs. 4.22 - 4.25, which critically includes the process $\pi^+ \to \mu^+ \nu_\mu$ followed by subsequent radiative muon decay). Summing these three spectral components weighted by their branching fractions as computed with the effective Lagrangians above gives the total spectrum from DM annihilation, $\dv{N}{E_\gamma}|_{\bar{\chi}\chi}$.

This spectrum can be combined with an assumed spatial distribution of DM to give the gamma-ray flux from DM annihilations in an astrophysical target subtending solid angle $\Delta\Omega$:
\begin{equation}
    \label{eq:flux_from_dm_ann}
    \left. \dv{\Phi}{E_\gamma} \right\rvert_{\bar{\chi}\chi}(E) = \frac{\Delta\Omega}{4\pi} \frac{\ev{\sigma v}_0}{2 f_\chi m_\chi^2} \bar{J} \left. \frac{dN}{dE_\gamma} \right\rvert_{\bar{\chi}\chi}(E).
\end{equation}
$\ev{\sigma v}_0$ is the thermally-averaged, present-day DM annihilation cross section in the target region. The $\bar{J}$ factor accounts for the amount of DM in the target, and depends on the square of the DM density $\rho$ since annihilations consume pairs of DM particles:
\begin{equation}
    \label{eq:j_factor_def}
    \bar{J} \equiv \frac{1}{\Delta\Omega} \int_{\Delta\Omega} \dd{\Omega} \int_{\u{LOS}} \dd{l} \rho(r(l,\psi))^2,
\end{equation}
$f_\chi$ accounts for the statistics of the DM, taking value $1$ when it is self-conjugate and $2$ otherwise. Since we assume the DM is a Dirac fermion, $f_\chi = 2$. In reality, a detector with finite energy resolution does not measure the flux in \cref{eq:flux_from_dm_ann} but rather the flux smoothed by an energy resolution function $R_\epsilon(E | E')$:
\begin{equation}
    \left. \dv{\Phi_\epsilon}{E_\gamma} \right\rvert_{\bar{\chi}\chi}(E) = \int \dd{E'} R_\epsilon(E | E') \left. \frac{d\Phi}{dE_\gamma} \right\rvert_{\bar{\chi}\chi}(E')
\end{equation}
The resolution function gives the probability a gamma ray with true energy $E'$ is reconstructed with energy $E$, and can generally be approximated as $\mathcal{N}\pqty{E | E', \epsilon(E') \, E'}$, with energy-dependent width parameter $\epsilon(E)$~\cite{Bringmann2009}.

Existing measurements of the gamma-ray flux from the Milky Way by the Imaging Compton Telescope (COMPTEL)~\cite{Strong:1998ck}, the Energetic Gamma Ray Experiment Telescope (EGRET)~\cite{Hunger:1997we}, the Fermi Large Area Telescope (Fermi-LAT)~\cite{FermiInstrument} and INTEGRAL/SPI~\cite{Bouchet:2011fn} constrain DM models. In lieu of attempting to model the astrophysical MeV gamma-ray background, we adopt the simple constraint-setting approach of requiring that the smoothed flux from DM annihilation integrated over each of a given detector's energy bins not exceed the measured flux plus twice the upper error in the bin: $\Phi_\epsilon^{(i)}|_{\bar{\chi}\chi} < \Phi^{(i)}|_\mathrm{obs} + 2 \sigma^{(i)}$, where $i$ indexes the bins. We have checked that setting a constraint using a $\chi^2$ test statistic yields similar results.

Several missions with capabilities in the MeV are in the proposal, planning, or construction phase. Here, we consider the following: AdEPT~\cite{adept}, AMEGO~\cite{amego}, All-Sky-ASTROGAM~\cite{as_astrogam}, GECCO~\cite{gecco_loi,alex_slides,Coogan:2021rez}, MAST~\cite{mast}, PANGU~\cite{pangu,pangu_aeff} and GRAMS~\cite{grams,grams_loi}. These are each characterized by an energy-dependent effective area $A_\mathrm{eff}(E)$ and energy resolution $\epsilon(E)$. For all telescopes we assume an observation time of 3 years, $T_\mathrm{obs} = \SI{9.5e7}{\second} = \SI{1095}{\day}$. 

Determining the reach of future telescopes requires adopting a background flux model $\dv{\Phi}{E_\gamma}|_\mathrm{bg}$ so the expected photon counts from DM annihilation and astrophysical processes in a target can be compared. A flux $\dv{\Phi}{E_\gamma}$ can be converted to photon counts at a detector over an energy range $[E_\mathrm{low}, E_\mathrm{high}]$ using
\begin{equation}
    \label{eq:count_from_dm_ann}
    N_\gamma = T_{\u{obs}} \int_{E_\u{min}}^{E_\u{max}} \dd{E} A_{\u{eff}} \, \dv{\Phi}{E_\gamma}(E)
\end{equation}
By substituting $\dv{\Phi_\epsilon}{E_\gamma}|_{\bar{\chi}\chi}$ and $\dv{\Phi}{E_\gamma}|_\mathrm{bg}$ into this equation we can construct the signal-to-noise ratio by maximizing over the energy range:
\begin{equation}
    \mathrm{SNR} \equiv \max_{E_\mathrm{min},\, E_\mathrm{max}} \frac{N_\gamma|_{\bar{\chi}\chi}}{\sqrt{N_\gamma|_{\mathrm{bg}}}}.
\end{equation}
We estimate that a DM model is discoverable if $\mathrm{SNR} > 5$.

Note that since the DM's mass is not identifiable under the background-only hypothesis, a full data analysis must compensate for the look-elsewhere effect using the methods of e.g. Ref.~\cite{Gross:2010qma}.\footnote{
    Detection significances must also be corrected slightly to account for the fact that $\ev{\sigma v}_0$ is non-negative, the so-called Chernoff correction to Wilks' theorem~\cite{10.2307/2236839}.
} We leave this more involved analysis for future work, but point out that except for the case where DM annihilates into pairs of photons the spectrum would be spread across multiple energy bins in a realistic analysis, reducing the trials factor correction to detection significances. The effect could also be mitigated by splitting the dataset (at the cost of a $\sqrt{2}$ reduction in SNR), and if a strong-enough annihilation signal is detected it could be confirmed by observing other targets such as M31 and Draco.

Since the $\bar{J}$ factor scales with the DM density squared, it is advantageous to select small, DM-rich targets for DM searches. We use the inner $\ang{10}\times \ang{10}$ region of the Milky Way as a target for the purposes of projecting the discovery reach of future telescopes. There is substantial uncertainty in modeling the baryonic and DM mass distributions in the Milky Way, which translates into factor of $10^{-1} - 10^1$ uncertainties in the $\bar{J}$ factor for the Galactic Center. We fix the DM distribution to the Einasto profile from Table III of Ref.~\cite{deSalas:2019pee}, with the parameters adjusted within their $1\sigma$ error bands to maximize $\bar{J}$. For consistency the constraints from existing gamma-ray observations are derived using this same profile. 

As a background model we adopt the one developed in Ref.~\cite{Bartels2017} specifically for searches in our target region. The model combines spectral components computed using the \texttt{GALPROP} cosmic ray propagation code~\cite{Strong:2001fu} as well as an analytic power-law component required to closely fit COMPTEL galactic center observations.

\begin{table}[tpb]
    \centering
    \begin{tabular}{ccccc}
        \toprule
        Detector & Latitude & Longitude & $\Delta\Omega$ (\si{\steradian}) & $\bar{J}$ (\si{\mega\eV^2.\centi\metre^{-5}.\steradian^{-1}}) \\
        \midrule
        COMPTEL~\cite{Kappadath:1993} & $|b| < \ang{20}$ & $|l| < \ang{60}$  & \num{1.433} & \num{1.04e30} \\
        EGRET~\cite{0004-637X-613-2-962}   & $\ang{20} < |b| < \ang{60}$ & $|l| < \ang{180}$ & \num{6.585} & \num{9.062e27} \\
        Fermi-LAT~\cite{0004-637X-750-1-3} & $\ang{8} < |b| < \ang{90}$ & $|l| < \ang{180}$ & \num{10.817} & \num{1.601e28} \\
        \midrule
        Planned telescopes & $|b| < \ang{5}$ & $|l| < \ang{5}$  & \num{3.042e-2} & \num{3.99e31} \\
        \bottomrule
    \end{tabular}
    \caption{Target regions and $\bar{J}$ factors for different detectors' measurements of the diffuse gamma-ray flux, as well as the target used to project the discovery reach of planned telescopes. All $\bar{J}$ factors were computed assuming the same Einasto profile (see text for details).}
    \label{tab:target_regions}
\end{table}

\subsection{Cosmic Microwave Background Constraints}%
\label{sub:cmb_constraints}

Dark matter annihilation around the time of CMB formation can inject ionizing particles into the photon-baryon plasma (see e.g. Ref.~\cite{Slatyer:2009yq}). The resulting changes in the residual ionization fraction and baryon temperature, modifying the CMB temperature and polarization power spectra, particularly at small scales. The changes depend on the energy per unit volume per unit time imparted to the plasma, quantified with the DM annihilation parameter
\begin{align}
    \label{eq:cmb_dm_ann_param}
    p_{\text{ann}} \equiv f_{\text{eff}}^\chi \cdot \frac{\ev{\sigma v}_{\text{CMB}}}{m_\chi}.
\end{align}
The thermal average is taken at the time of CMB formation. In the case of $s$-wave annihilation, $\ev{\sigma v}_{\text{CMB}}$ is equal to the present-day thermally averaged annihilation cross section, $\ev{\sigma v}_0$. For $p$-wave annihilation, $\ev{\sigma v}_{\text{CMB}} = \ev{\sigma v}_0 \, (v_{\text{CMB}} / v_0)^2$, where we take the present-day DM velocity to be $v_0 = \SI{220}{\kilo\metre\per\second}$ in the Milky Way. The DM velocity at recombination depends on the kinetic decoupling temperature. While the DM velocity is thermal before kinetic decoupling, it redshifts more quickly afterwards, giving \cite{Essig:2013goa}
\begin{align}
    \label{eq:v_cmb}
    v_{\text{CMB}} & = \sqrt{3 T_\chi / m_\chi} \approx 2 \times 10^{-4} \left( \frac{T_\gamma}{1~\text{eV}} \right) \left( \frac{1~\text{MeV}}{m_\chi} \right) \left( \frac{10^{-4}}{x_{\text{kd}}} \right)^{1/2},
\end{align}
where $x_{\text{kd}} \equiv T_\chi / m_\chi \approx 10^{-4} {-} 10^{-6}$ and we take $T_\gamma = \SI{0.235}{\eV}$. This means that while the CMB bounds on $\ev{\sigma v}_0$ are quite stringent for $s$-wave annihilation, they are much weaker for $p$-wave annihilation since $v_\mathrm{CMB}$ is much smaller than $v_0$. The quantity $f_{\text{eff}}^\chi$ encapsulates how efficiently DM annihilations inject energy into the plasma. It is obtained by integrating the $e^\pm$ and photon spectra per DM annihilation weighted by energy injection efficiency factors~\cite{PhysRevD.93.023527}. This quantity can be computed in \texttt{hazma} (see sec. 8 of Ref.~\cite{hazma}).

In the constraint plots below, we use the most recent Planck constraint on the DM annihilation parameter \cite{PlanckConstraint}, $p_{\text{ann}} < \SI{4.1e-31}{cm^3.s^{-1}.\mega\eV^{-1}}$.

\subsection{Higgs portal model constraints}\label{sec:scalarpheno}

There are numerous terrestrial constraints on our Higgs portal model (see e.g. Ref.~\cite{Krnjaic:2015mbs}), and in particular the coupling $\sin\theta$. The most important here are the upper limit on the invisible Higgs branching fraction, bounds on several rare visible and invisible meson decays processes and limits on light particle production at beam dumps. 

For the first process, the invisible Higgs branching fraction can be computed from the microscopic interaction Lagrangian and compared with current constraint of $\mathrm{Br}(h \to \bar{\chi} \chi) < 0.19$ from the CMS detector at the Large Hadron Collider~\cite{Sirunyan:2018owy}. In our analysis the DM mass is much smaller than the Higgs mass, so this constraint is independent of $m_\chi$.

The scalar mediator gives a contribution to flavor-changing neutral currents (FCNCs) that cause rare $B$ and $K$-meson decays. These proceed via $B \to K \, S$ and $K \to \pi \, S$, followed by subsequent decay of the mediator into leptons or DM particles. We follow the analysis in Appendices A and C of Ref.~\cite{Dolan:2014ska} to compute constraints on our model from Belle measurements of $\mathrm{Br}(B^+ \to K^+ \ell^+ \ell^-)$~\cite{PhysRevLett.103.171801}, KTeV's upper limits on $\mathrm{Br}(K_L \to \pi^0 \ell^+ \ell^-)$~\cite{AlaviHarati:2003mr,PhysRevLett.84.5279}, BaBar's upper limit on $\mathrm{Br}(B^+ \to K^+ \bar{\nu}\nu)$~\cite{PhysRevD.87.112005} and E949's measurement of $\mathrm{Br}(K^+ \to \pi^+ \bar{\nu}\nu)$~\cite{PhysRevD.77.052003,PhysRevD.79.092004}.

The CHARM beam dump searched for electrons, muons and photons produced by the decays of light particles created in collisions between a 400 GeV proton beam and a copper target~\cite{Bergsma:1985qz}. Since no such decays were detected, the number of decays in the detector volume is constrained to be less than 2.3 at the \SI{90}{\percent} confidence level~\cite{Clarke:2013aya}. We follow the analysis in Ref.~\cite{Clarke:2013aya} to impose this constraint.

The complementarity between these terrestrial constraints and indirect detection constraints and projections in the $(m_\chi, \ev{\sigma v})$ plane depends on whether the DM annihilates invisibly or visibly. In the first case, corresponding to a mediator mass smaller than the dark matter mass, the cross section $\ev{\sigma v}$ is proportional to $g_{S\chi}^4$: $\sin \theta$ plays no role in indirect detection. The only requirement in this case is that there is \emph{some} $\sin\theta$ value consistent with terrestrial observations, and that it is large enough that the mediator's decay length is below the parsec scale. This is the case for all DM masses we will consider.

\begin{figure}
    \centering
    \includegraphics[width=\textwidth]{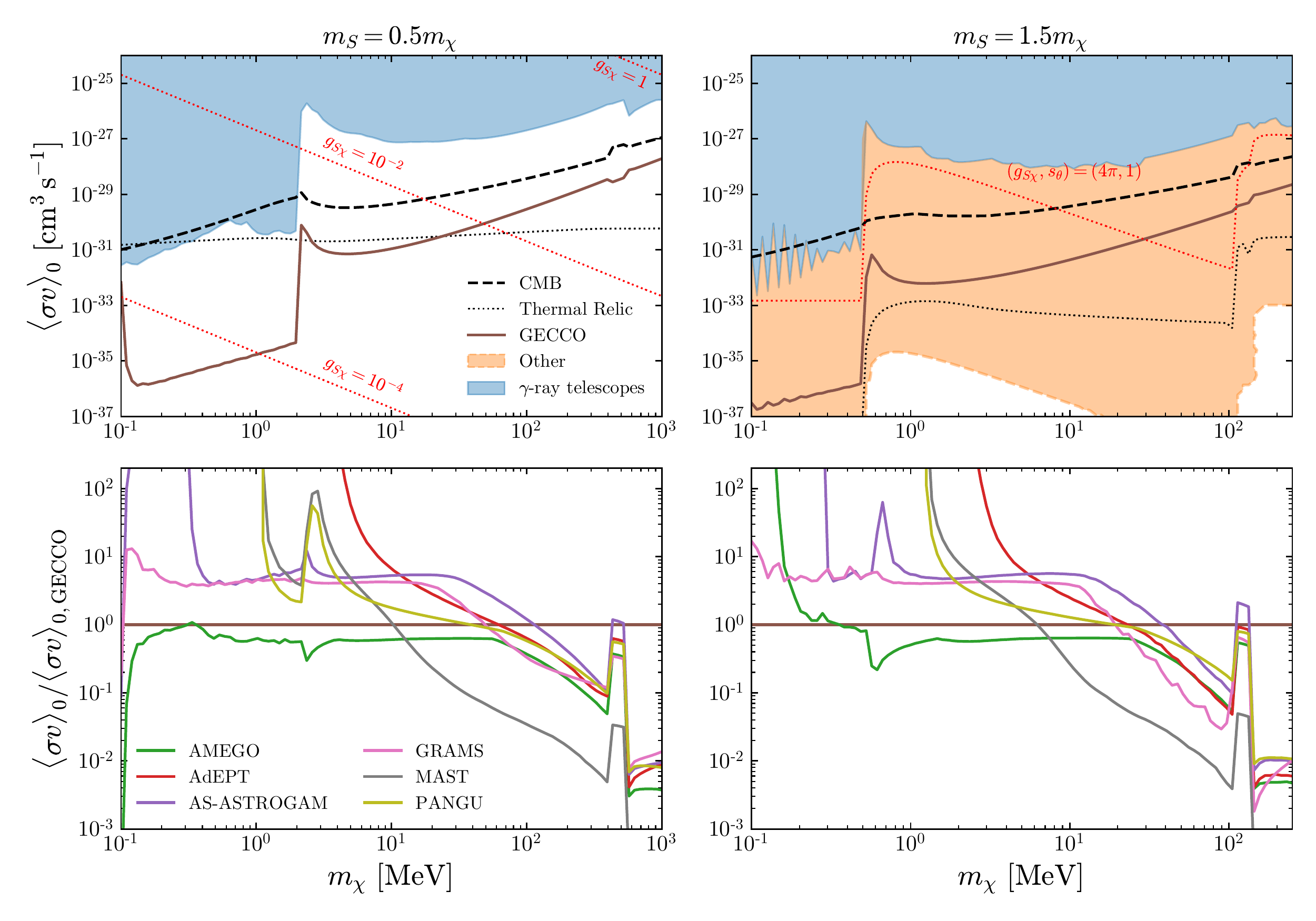}
    \caption{Prospects for future gamma-ray telescopes, and phenomenological constraints, for the Higgs portal case; the left panels assume a mediator mass half the mass of the DM, while the right panels assume a mediator mass equal to 1.5 times the DM mass. The lower panels show the future telescope prospects normalized to GECCO's (a smaller value on the $y$ axis corresponds to a {\em tighter} constraining capability). In computing the CMB constraint we conservatively assume a kinetic decoupling temperature of $10^{-6} \, m_\chi$.}
    \label{fig:hp_constraints}
\end{figure}

If instead the DM annihilates into SM final states (i.e. the mediator mass is larger than the DM mass), the annihilation cross section scales as $g_{S\chi}^2 \, \sin^2\theta \, y^2$, where $y \ll 1$ is the Yukawa coupling of the final state particles. The Yukawa strongly suppresses the indirect detection signal, and large values of $g_{S\chi}$ and $\sin\theta$ are required to give detectable signals (see the dashed red line in the right panel of \cref{fig:hp_constraints}). We translate terrestrial constraints into the $(m_\chi, \ev{\sigma v})$ plane as follows. For fixed $m_\chi$, each point in this plane defines a range of $\sin\theta$ and corresponding $g_{S\chi}$ values. The largest-possible value of $\sin\theta$ is $1$ and the smallest value corresponds to $g_{S\chi} \sim 4 \pi$. We thus check at each point whether there is \emph{some} value of $(\sin\theta,\, g_{S\chi})$ in this range that is consistent with terrestrial constraints. If there is not, that point is excluded, and we highlight it in orange in our final constraint plots (\cref{fig:hp_constraints}).

\subsection{Dark photon model constraints}\label{sec:vectorpheno}

\begin{figure}
    \centering
    \includegraphics[width=\textwidth]{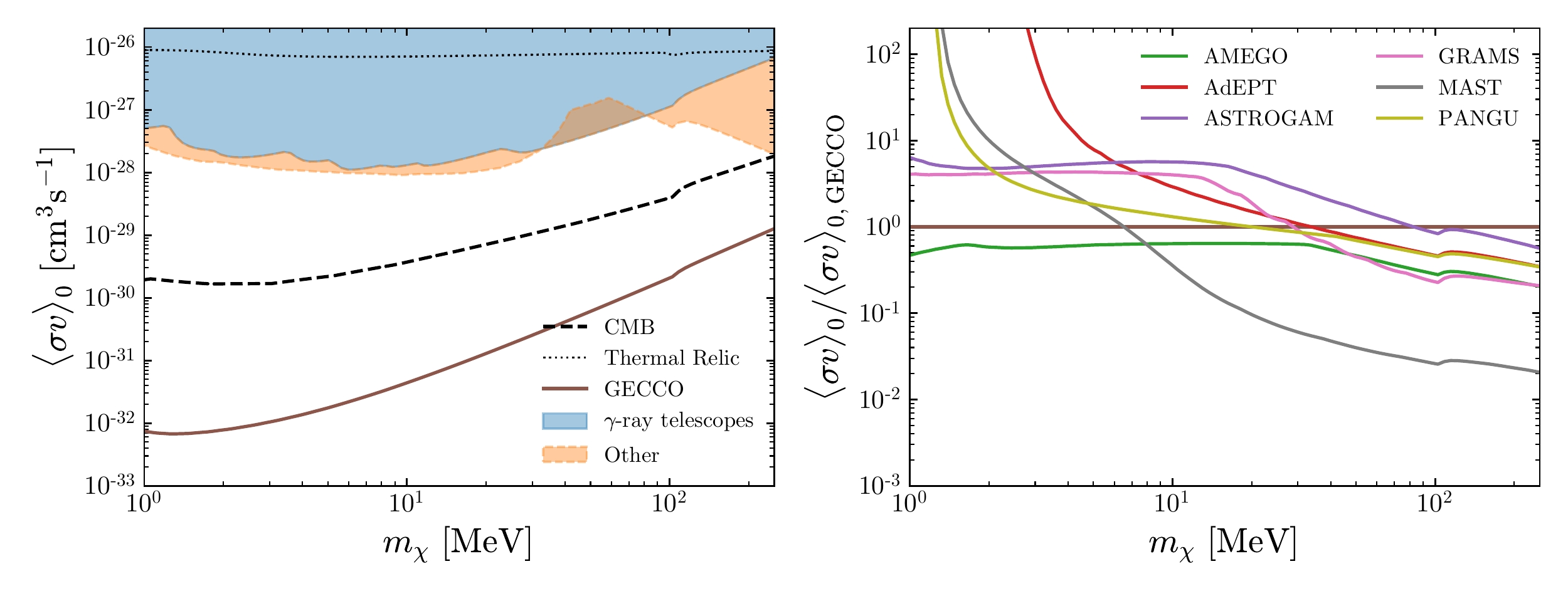}
    \caption{Vector mediator constraints, for a mediator mass \(m_{V}=3 m_{\chi} \). The right panel shows the ratio of the projected constraints from a given future telescope relative to those for GECCO (shown with a brown line in the left panel).}
    \label{fig:km_constraints}
\end{figure}

We choose here to focus on the regime where the mediator is heavier than the dark matter mass, and assume \(3 m_{\chi} = m_{V}\). This choice enables us to re-use previously-calculated  constraints from non-astrophysical experiments. The strongest constraints on dark photon models for the masses we are interested in come from the \(B\)-factory BaBar~\cite{Lees_2017} and beam-dump experiments such as LSND~\cite{Aguilar_2001}. Studies using the datasets of these experiments were able to constraint the dark photon model by searching for dark photon production and subsequent decay into dark matter (see, for example Ref.~\cite{batell2009exploring, lees2017search, 10.1093/ptep/ptz106}); in the case of BaBar, the relevant process is \(\Upsilon(2S),\Upsilon(3S)\to \gamma + V \to \gamma + \mathrm{invisible}\), while the relevant process for LSND is \(\pi^{0}\to\gamma + V \to \gamma + \mathrm{invisible}\). We adapt here the constraints computed in Ref.~\cite{aakesson2018light,10.1093/ptep/ptz106} (see Fig.(201) of Ref.~\cite{10.1093/ptep/ptz106} for the constraints and the text and references therein for details). These references take $\alpha_D = g_{V\chi}^2/4\pi = 0.5$ and we do the same in order to consistently compare their constraints with our gamma-ray constraints. This value is conservative for the phenomenological constraints. Lowering $\alpha_D$ will strengthen the phenomenological constraints (for example, taking $\alpha_D = 10^{-3}$ will strengthen the constraints by an order of magnitude; see Ref.~\cite{aakesson2018light}). However, the gamma-ray constraints will be unaffected.

\subsection{Results}%
\label{sub:results}

\Cref{fig:hp_constraints} shows the current and projected limits on the scalar mediator (Higgs portal) model. In the left panels, the DM pair annihilates preferentially to the scalar mediator, which, in turn, decays to kinematically-available final states according to the corresponding Yukawa coupling. A prominent feature of the gamma-ray constraints lines (including from existing telescopes) appears at the electron threshold, corresponding to a dark matter mass being four times the electron mass (since the mediator's mass is half that of the dark matter, and to decay into electron-positron pairs its mass must be twice the mass of the electron). The other visible features, at higher DM masses, correspond to the muon and pion thresholds.

Since the mixing angle $\theta$ is virtually unconstrained in the invisible decay case (left panels), phenomenological constraints are weaker, in general, than existing gamma-ray constraints. Also, notice that because of the velocity suppression of the pair-annihilation cross section in this model, CMB limits are weak compared to the expected cross section from thermal production.

The anticipated performance of future MeV gamma-ray telescopes is shown with the brown line for GECCO in the upper panel, and in the lower panel for all other future telescopes under consideration, relative to GECCO (a smaller ratio corresponding to tighter constraints). While the relative sensitivity of telescopes such as GRAMS, All-Sky ASTROGAM, and AMEGO are within a factor of a few for several decades in DM mass, the sudden appearance of muons in the annihilation final state, and the corresponding final-state radiation, samples in a non-trivial way the effective area of telescopes as a function of energy in the DM mass range between the muon and pion threshold. This produces the sudden sensitivity decrease for virtually all telescopes compared to GECCO in that DM mass range, visible as a box-shaped feature.

Broadly, we find that as soon as the mediator decay to $e^+e^-$ opens up, future MeV telescopes will definitely be sensitive to thermally-produced MeV dark matter for light scalar mediators where the dark matter pair annihilates preferentially invisibly.

The situation is markedly different for heavier scalar mediator. Generally, most parameter space is, in this case, ruled out by phenomenological constraints. While in principle future MeV telescopes will be sensitive to this case as well (as shown by the red-dotted line being {\em above} the brown line in the top, right panel), it will be critical for the DM mass to exceed the electron threshold, and/or the muon threshold. Again, though, future MeV telescope will outperform CMB constraints as well as, naturally, existing gamma-ray telescopes.

The case of dark matter annihilation via kinetically-mixed vector mediators weighing {\em more} than the dark matter is shown in \cref{fig:km_constraints}. Current gamma-ray telescopes do not provide any meaningful constraint, given current phenomenological constraints, as shown by the orange shaded region extending beyond the blue shaded region in the left panel. Additionally, for all but the heaviest masses CMB constraints are the strongest for this case, where the pair-annihilation cross section is {\em not} velocity suppressed. Nonetheless, our analysis indicates that future MeV gamma-ray telescopes such as GECCO (whose projected sensitivity is shown with a brown line in the left panel) will out-perform CMB constraints by between 1 and 2 orders of magnitude, thus being slated to probe meaningful portions of open parameter space.

Compared with other gamma-ray telescopes, GECCO's performance is similar, albeit slightly inferior, to AMEGO's, but better than All-Sky ASTROGAM up to 100 MeV masses, and better than even MAST up to almost 10 MeV. Again, smaller cross section ratios (shown in the $y$-axis) indicate more constraining power.

%% file: sections/conclusions.tex
This study aimed at filling a few gaps in the reliable analysis of indirect detection prospects for annihilating dark matter in the MeV mass range. The key analytical results we presented were the matching of parton-level interactions between Higgs portal and dark photon mediators onto meson-level ones, which were obtained by means of chiral perturbation theory. We clarified the range of validity of our leading-order chiral perturbation theory treatment as a function of the dark matter and mediator mass. We then compared current constraints from the CMB, terrestrial experiments and existing and past gamma-ray telescopes with the anticipated performance of planned telescopes. Our gamma-ray constraints/projections and CMB ones were derived by computing annihilation branching ratios using the meson-level interactions we derived, and applying our previous analysis of the final-state radiation and radiative decay spectra for light meson and lepton final states~\cite{hazma}.

We focused on invisibly- and visibly-annihilating Higgs-portal mediator models. For the former, future gamma-ray telescopes will explore large swaths of yet-unexplored parameter space (including that preferred by thermal dark matter production), while for the latter, typically the existing phenomenological constraints will prevent exploring new models. For the dark photon mediator case, we found that new MeV gamma-ray telescopes will enhance the constraints on the dark matter self-annihilation cross section by 1 to 2 orders of magnitude beyond the values currently constrained by CMB observations.

While the landscape of WIMP searches in the traditional \emph{GeV-TeV} range appears to offer faltering returns for dark matter discovery, our study illustrates how future \emph{MeV} gamma-ray detectors provide very promising prospects over several orders of magnitude in the dark matter mass. With improved observational tools, improved theoretical tools are in order. This paper addressed some of the shortcomings in the current understanding of how to reliably compute gamma-ray spectra, and elucidated how, with the technique of chiral perturbation theory, the discovery of an anomalous gamma-ray spectrum might be tied to a parton-level Lagrangian, potentially offering important guidance to illuminating the nature of physics beyond the Standard Model of particle physics.